\documentclass[prl,showpacs,showkeys,superscriptaddress,amsfonts,amsmath,twocolumn]{revtex4}
\usepackage{graphicx} 
\usepackage{subfigure}

\newcommand{\ket}[1]{|#1\rangle}
\newcommand{\bra}[1]{\langle #1|}

\begin{document}

 \title{Non-cyclic Geometric Phase due to Spatial Evolution in
a Neutron Interferometer}

\affiliation{Atominstitut der {\"{O}}sterreichischen Universit{\"{a}}ten,
Stadionallee 2, A-1020 Vienna, Austria}
\affiliation{Institut Laue Langevin, Bo\^ite Postale 156, F-38042 Grenoble
Cedex 9, France}
\author{Stefan Filipp}
\email{sfilipp@ati.ac.at}
\affiliation{Atominstitut der {\"{O}}sterreichischen Universit{\"{a}}ten,
Stadionallee 2, A-1020 Vienna, Austria}
\affiliation{Institut Laue Langevin, Bo\^ite Postale 156, F-38042 Grenoble
Cedex 9, France}
\author{Yuji Hasegawa}
\email{hasegawa@ati.ac.at}
\affiliation{Atominstitut der {\"{O}}sterreichischen Universit{\"{a}}ten,
Stadionallee 2, A-1020 Vienna, Austria}
\author{Rudolf Loidl}
\affiliation{Institut Laue Langevin, Bo\^ite Postale 156, F-38042 Grenoble
Cedex 9, France}
\author{Helmut Rauch}
\affiliation{Atominstitut der {\"{O}}sterreichischen Universit{\"{a}}ten,
Stadionallee 2, A-1020 Vienna, Austria}

\begin{abstract}
We present a split-beam neutron interferometric experiment to test
the non-cyclic geometric phase tied to the spatial evolution of the
system: the subjacent two-dimensional Hilbert space is spanned
 by the two possible paths in the
interferometer and the evolution of the state is controlled by
phase shifters and absorbers. A related experiment was reported
previously by Hasegawa \emph{et al.} [Phys. Rev. A {\bf 53}, 2486 (1996)] to verify the
\emph{cyclic} spatial geometric phase. The interpretation of this
experiment, namely to ascribe a geometric phase to this particular
state evolution, has met severe criticism from Wagh [Phys. Rev. A
{\bf 59}, 1715 (1999)]. The
extension to a \emph{non-cyclic} evolution manifests the correctness
of the interpretation of the previous experiment by means of an
explicit calculation of the non-cyclic geometric phase in terms of
paths on the Bloch-sphere.
 \end{abstract}


\maketitle

Reported already by Pancharatnam \cite{pancharatnam56} in the
1950s a vast amount of intellectual work has been put into the
investigation of geometric phases. In particular, Berry showed in
1984 \cite{berry84} that a geometric phase arises for the
adiabatic evolution of a quantum mechanical state which triggered
renewed interest in this topic. The evolution of a system
returning to its initial state causes an additional phase factor
connected only to the path transversed in state space. There
have been several extensions in various directions
\cite{wilczek-zee84,aharonov-anandan87,samuel-bhandari88,mukunda-simon93,pati95,manini-pistolesi00}
for pure states, but also for the mixed state case
\cite{uhlmann86,sjoqvist00,filipp-sjoqvist03}. Besides these
theoretical work numerous experiments have been performed to
verify geometric phases using various types of quantum mechanical
systems, e.~g. polarized photons \cite{tomita-chiao86} or NMR
\cite{du03}. In addition, neutron interferometry has been established as a
particularly suitable tool to study basic principles of quantum
mechanics \cite{rauch00,rauch02,hasegawa03} providing 
explicit demonstrations 
\cite{bitter-dubbers87, wagh98, hasegawa96, hasegawa01} and facilitating
further studies \cite{bertlmann04} of geometric phenomena.

There is no reason to consider only inherent quantum properties
like spin and polarization for the emergence of a geometric phase;
equally well one can consider a subspace of the momentum-space of
a particle and its geometry. On this issue some authors of the
present article  performed an experiment to test the spatial
geometric phase \cite{hasegawa96} .
The results are fully
consistent with the values predicted by theory, however, there is
an ambiguity in the interpretation as pointed out by Wagh
\cite{wagh99}. He concludes that in this setup the phase picked up
by a state during its evolution is merely a U(1) phase factor
stemming from the dynamics of the system and not due to the
geometric nature of the subjacent Hilbert space. 

 In this paper we generalize the idea of the experiment in
\cite{hasegawa96} to resolve the ambiguity in the interpretation
of this antecedent neutron interferometry experiment. There the
geometric phase has been measured for a $2\pi$ (cyclic) rotation
of the Bloch-vector representing the path state of the neutron. In
order to deny Wagh's criticism we have now measured the geometric
phase for a rotation by an angle in the intervall $[0,2\pi]$
(non-cyclic) and -- to show the applicability of the geometric
phase concept -- we have devised the path of the state vector on
the Bloch-sphere to calculate the corresponding surface area
enclosed by the evolution path. In the theory this surface area
is proportional to the geometric phase, which has been determined
experimentally to confirm the validity of our considerations
and therefore the proper
 interpretation in terms of a geometric phase.

 \begin{figure}[htbp]
  \centering
  \includegraphics[width=80mm]{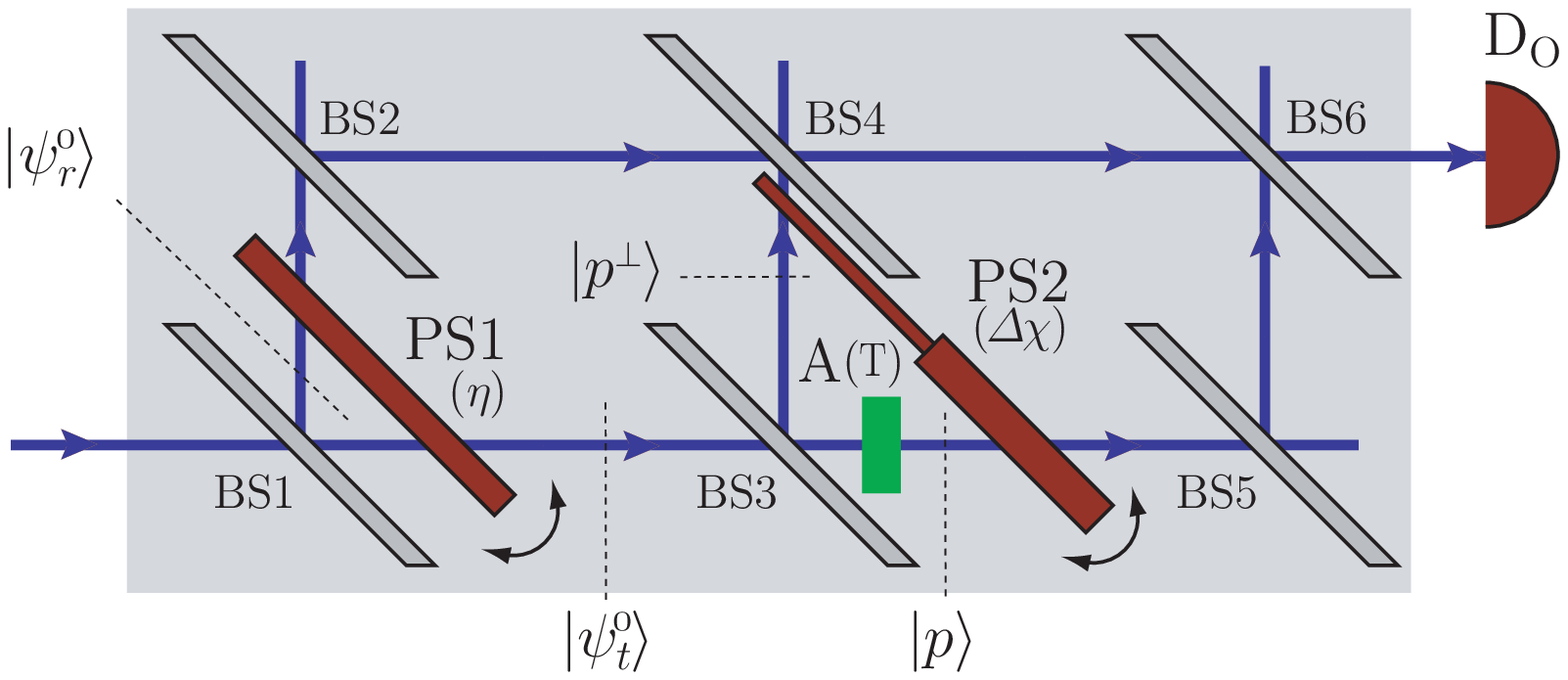}
  \caption{Experimental setup utilizing a double-loop perfect-crystal neutron interferometer:
 One loop is used for the state manipulation with a phase shifter (PS2) together with a beam attenuator(A)
 and the other one provides a reference beam with adjustable
 phase by use of the phase shifter (PS1).}
  \label{2004-splitbeam-prl-f1}
\end{figure}

For testing the spatial geometric phase we use a double-loop
interferometer (cf. Fig. \ref{2004-splitbeam-prl-f1}), where the
incident (unpolarized) neutron beam $\ket{\psi}$ is split at the
beam splitter BS1 into a reflected beam $\ket{\psi_{r}^0}$ and a
transmitted beam $\ket{\psi_t^0}$.

The reflected beam $\ket{\psi_r^0}$ is used as a reference with
adjustable relative phase $\eta$ to $\ket{\psi_t^0}$ due to the
phase shifter PS1. The latter beam is defined to be in the state
$\ket{\psi_t^0} \equiv \ket{p}$ before the beam splitter BS3, where
$\ket{p}$ is the eigenstate to the operator $P_p \equiv \ket{p}\bra{p}$
measuring the path. Behind BS3 there are two possible orthogonal
paths $\ket{p}$ and $\ket{p^\perp}$ spanning a two-dimensional
Hilbert space, where $\ket{p}$ denotes the state of the
transmitted beam and $\ket{p^\perp}$ the state of the reflected
beam, respectively. Having a 50:50 beam splitter
$\ket{\psi_t^0}$ is transformed into a superposition
 of the basis vectors $\ket{p}$ and
$\ket{p^\perp}$: $\ket{\psi_t^0} \mapsto \ket{q}\equiv (\ket{p} +
\ket{p^\perp})/\sqrt{2}$.  The corresponding projection operator $P_q
\equiv \ket{q}\bra{q}=(1 + \ket{p}\bra{p^\perp} +
\ket{p^\perp}\bra{p})/2$ (and also $P_{q^\perp} = 1 - P_q =
\ket{q^\perp}\bra{q^\perp}$) measures the interference instead of
the paths.

The transmitted beam $\ket{\psi_t^0}$ is subjected to further
evolution in the second loop of the interferometer by use of beam
splitters (BS4, BS5 and BS6), an absorber (A) with transmission
coefficient $T$ and a phase shifter (PS2) generating a phase shift
of $e^{i\chi_1}$ on the upper ($\ket{p^\perp}$) and $e^{i\chi_2}$
on the lower beam path ($\ket{p}$), respectively, yielding the
final state $\ket{\psi_{t}}$. Thus, the evolution causing the
spatial geometric phase can be written as

\begin{eqnarray}
  \label{eq:evolution}
  \ket{\psi_t^0} & \xrightarrow{\text{BS3}}&
  \frac{1}{\sqrt{2}}(\ket{p^\perp}+\ket{p}) \xrightarrow{\text{A}}
    \frac{1}{\sqrt{2}}(\ket{p^\perp}+\sqrt{T}\ket{p})\\
    &\xrightarrow{\text{PS2}}&\frac{1}{\sqrt{2}}(e^{i\chi_1}\ket{p^\perp}+\sqrt{T}e^{i\chi_2}\ket{p}) \equiv \ket{\psi_{t}}\nonumber.
\end{eqnarray}

The transformation of the reference beam $\ket{\psi_r^0}$ is given
by $\ket{\psi_r^0} \mapsto e^{i\eta}\ket{\psi_r^0} \mapsto e^{i\eta}\ket{p^\perp}$, which follows
from the fact that the path of $\ket{\psi_r^0}$ coincides with the
path of the beam reflected at BS3 labeled by $\ket{p^\perp}$.

In the last step $\ket{\psi_{t}}$ and the reference beam are
recombined at BS6 and detected in the forward beam at the detector
D$_O$.
This recombination
 can be described by application of the interference projection
operator $P_{q}=\ket{q}\bra{q}$ to $\ket{\psi_{t}}$ as well as to
$\ket{\psi_r^0}$:
  \begin{eqnarray}
        \label{eq:detection}
    \ket{\psi_{t}^\prime}&\equiv&P_{q}{\psi_{t}}=K(e^{i\chi_1}+\sqrt{T}e^{i\chi_2})\ket{q}\nonumber\\
    e^{i\eta}\ket{\psi_{r}^\prime}&\equiv&P_{q}e^{i\eta}{\ket{\psi_r^0}} =Ke^{i\eta}\ket{q},
  \end{eqnarray}
where $K$ is some scaling constant.

The intensity $I$ measured in the detector $D_O$ is proportional
to the modulus squared of the superposition $\ket{\psi_t^\prime}+
e^{i\eta}\ket{\psi_{r}^\prime}$:
\begin{eqnarray}
  \label{eq:interference}
  I&\propto&\big|\ket{\psi_t^\prime}+
e^{i\eta}\ket{\psi_{r}^\prime}|^2 = \langle \psi_{r}^\prime
  | \psi_{r}^\prime \rangle + \langle \psi_{t}^\prime | \psi_{t}^\prime \rangle +\nonumber\\
&& + 2 |\langle \psi_{r}^\prime | \psi_{t}^\prime \rangle |
\cos( \eta -\Phi)
\end{eqnarray}
with $\Phi\equiv \arg \langle \psi_{r}^\prime | \psi_{t}^\prime
\rangle$.
Explicitly, using Eq. (\ref{eq:detection}) we obtain
\begin{eqnarray}
  \label{eq:phase}
  \Phi &=& \frac{\chi_1 + \chi_2}{2} + \arg \left(e^{-i\frac{\Delta\chi}{2}} + \sqrt{T}e^{i\frac{\Delta\chi}{2}}\right)\nonumber\\
  &=&  \frac{\chi_1+ \chi_2}{2} -
  \arctan\left[\tan\left(\frac{\Delta\chi}{2}\right)\left(\frac{1-\sqrt{T}}{1+\sqrt{T}}\right)\right],
\end{eqnarray}
where $\Delta\chi\equiv\chi_2 - \chi_1$. By varying $\eta$ we can read off $\Phi$ as a shift of the
interference pattern.

For our purposes a double loop interferometer is inevitable, since
we measure the phase shift generated in one interferometer loop
relative to the reference beam, in contrast to a phase difference between
two paths measured in usual interferometric
setups. Here, the relative phase difference $\Phi$
between $\ket{\psi_t'}$ and $\ket{\psi_r'}$ provides information about the evolution of the
state $\ket{\psi_t^0}$ in state space.
The
geometric phase $\Phi_g$ is defined as
$\Phi_g \equiv \Phi - \Phi_d$  \cite{mukunda-simon93},
where $\Phi_d$ denotes the dynamical part. In our setup
$\Phi_d$ stems from the phase shifter PS2 and is
given by a sum of the phase shifts $\chi_1$ and $\chi_2$ weighted
with the transmission coefficicient $T$ \cite{hasegawa96, wagh99},
$\Phi_d =(\chi_1 + T\chi_2)/(1+T)$. It vanishes by an
appropriate choice of positive and negative phase shifts in accordance
with the transmission, i.~e. $\Phi_d
= 0$ for $-\chi_1/\chi_2 = T$.

\begin{figure}[htbp]
  \centering
\subfigure[Cyclic
evolution]{\label{2004-splitbeam-prl-f2a}\includegraphics[width=40mm]{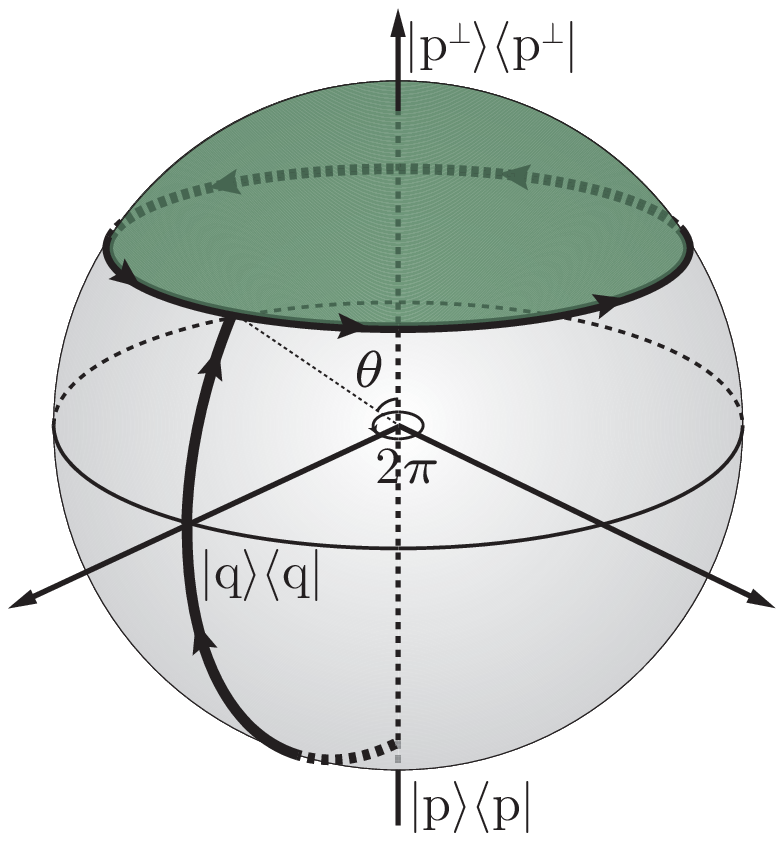}}
  \subfigure[Non-cyclic
  evolution]{\label{2004-splitbeam-prl-f2b}\includegraphics[width=40mm]{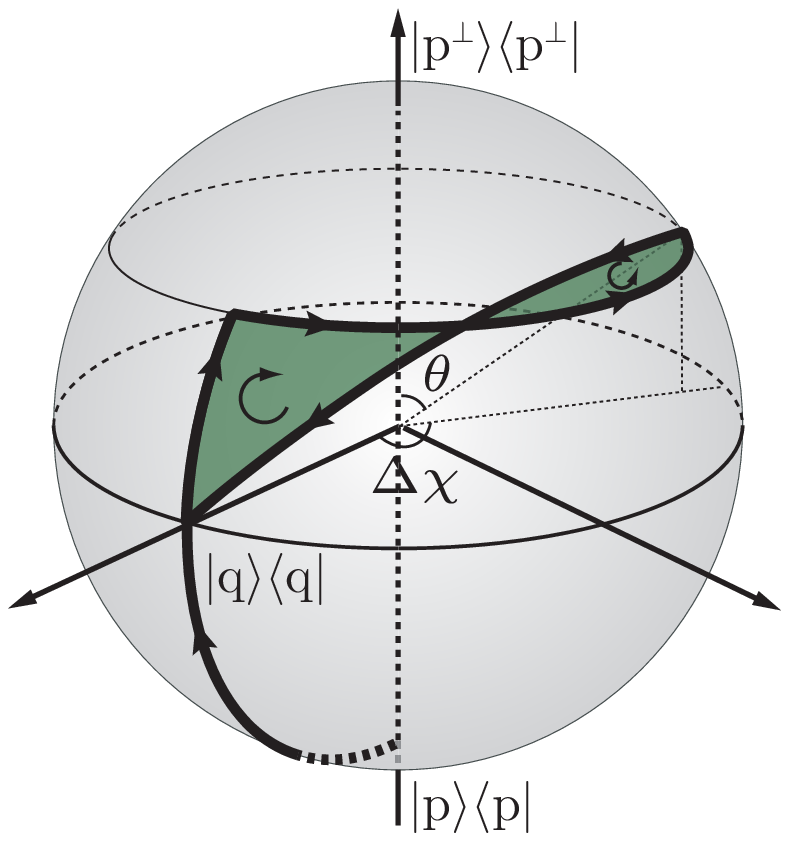}}
  \caption{Paths on the Bloch-sphere corresponding to the evolution of
    the state $\ket{\psi_t^0}$ (a) in the  cyclic and (b) non-cyclic case.
    The enclosed solid angle as seen from the origin of the sphere is proportional
    to the geometric phase observed in the experiment.}
  \label{2004-splitbeam-prl-f2}
\end{figure}

Note, that the same evolution can also be implemented in spin space by
thinking of polarizers instead of beam splitters, and
magnetic fields instead of absorber and second phase shifter PS2. The
phase shift for such a setup differs from $\Phi$ in
Eq. (\ref{eq:phase}) merely by a purely dynamical contribution that
is compensated in our experiment.

The result from Eq. (\ref{eq:phase}) can also be obtained  by
purely geometric considerations. Since we are dealing with a
two-level system corresponding to the possible paths of
$\ket{\psi_t^0}$ in the second loop of the interferometer the state
space is equivalent to a sphere in $\mathbb{R}^3$, known as the
Bloch-sphere \cite{mittelstaedt87, busch87, hasegawa94}. From theory we know
that the geometric phase $\Phi_g$ 
is given by the (oriented) surface area enclosed by the path of
the state vector on the Bloch-sphere and is
proportional to the enclosed solid angle as
seen from the origin of the sphere.

To
each point on the sphere there is a corresponding projection
operator. 
As basis we choose 
$\ket{p^\perp}\bra{p^\perp}$ and $\ket{p}\bra{p}$ represented as the
north and the south pole of the sphere, respectively
(Fig. \ref{2004-splitbeam-prl-f2}). At the beam splitter BS3 the state $\ket{\psi_t^0}$ originating from
the point $\ket{p}\bra{p}$ is projected to an equal superposition of
upper path and lower path depicted as a geodesic from the south
pole to the equatorial line on the Bloch-sphere \footnote{The
particular point on the equator is arbitrary due to the arbitrary
choice of the phases of the basis vectors.}.

The absorber with transmittivity $T=\tan^2\theta/2$, $\theta\in
[0,\pi/2]$, changes the weights of the superposed basis states
$\ket{p}$ and $\ket{p^\perp}$. The resulting state is encoded as a
point on the geodesic from the north pole to the equatorial line. In
particular for no absorption ($\theta=\pi/2$ or $T=1$ ) the state stays
on the equator. By inserting a beam block ($\theta=0$ or $T=0$) there
is no contribution from $\ket{p}$ so that the state is pinned onto the
north pole.

The phase shifter PS2 induces a relative phase shift
between the superposing states of $\Delta\chi = \chi_2-\chi_1$.
This corresponds to an evolution along a circle of latitude on the
Bloch-sphere with periodicity $2\pi$.
The recombination at BS5 followed by the detection of the forward
beam in D$_\text{O}$ is represented as a projection to the
starting point on the equatorial line, i.~e. we have to close the
curve associated with the evolution of the state by a geodesic to
the point $\ket{q}\bra{q}$ in accordance to the results in \cite{samuel-bhandari88}.

This evolution path is depicted in Figs. \ref{2004-splitbeam-prl-f2a}
and \ref{2004-splitbeam-prl-f2b} for cyclic and non-cyclic evolution, respectively.
For a relative phase difference greater than $\pi/2$ we have to
take the direction of the loops into account: In Fig. \ref{2004-splitbeam-prl-f2b} the first loop is
transversed clockwise, whereas the second loop is transversed
counter-clockwise yielding a positive or negative contribution to the
geometric phase, respectively.

With this representation we can numerically calculate the solid angle
$\Omega$ enclosed by the transversed path on the Bloch-sphere. The
resuls obtained in this way for $\Phi_g = -\Omega/2$ \cite{berry84}
 are equal to the results based on Eq. (\ref{eq:phase}). 
This substantiates the emergence
of a geometric phase in this type of experiment contrary to other
claims \cite{wagh99}.


As for the experimental demonstration we have used the double-loop
perfect-crystal-interferometer installed at the S18 beamline at
the high-flux reactor ILL, Grenoble \cite{zawisky02}. A schematic
view of the setup is shown in Fig. \ref{2004-splitbeam-prl-f1}.
Before falling onto the skew-symmetric interferometer the incident
neutron beam is collimated and monochromatized by the 220-Bragg
reflection of a Si perfect crytal monochromator placed in the
thermal neutron guide H25. The wavelength is tuned to give a mean
value of $\lambda_0 = 2.715 \text{\AA}$.
 The beam cross-section is confined to
$5\times 5 \text{mm}^2$ and by use of an isothermal box enclosing the
interferometer  thermal environmental isolation is achieved. As
phase shifters parallel sided Al plates are used. In fact,
a 5mm-thick plate
  is taken for the first phase shifter (PS1) inserted in the former loop
 and plates of different thickness ($d_1=0.5$mm and $d_2=4.1$mm) are used as
 the second phase shifter (PS2). 

The different thicknesses together with a specific choice of the absorber (A)
are to eliminate a phase of unwanted dynamical origin. 
In each beam a positive phase shift $\chi_{1,2}\propto d_{1,2}$ is
induced by PS2 \cite{rauch00}. By a rotation of this phase shifter
through a (small) angle $\gamma$ about an axis perpendicular to
the interferometer $\chi_1$ and
$\chi_2$ change with opposite sign, i.~e., $\Delta\chi_1\propto
- d_1\gamma$,
while $\Delta\chi_2 \propto d_2\gamma$. For the relative
phase shift $\Delta\chi=\chi_2-\chi_1$ between the two paths we have 
$\Delta\chi - \chi_0 \propto (d_2+d_1) \gamma$, where the constant $\chi_0$ --
determined by the initial position of PS2 -- has been adjusted to
$\chi_0 = 2n\pi$, $n$ integer, and  can thus be neglected.

Furthermore, we have intended to set the transmission coefficients
$T_j$ of each beam after PS2 and A as $T_2/T_1 =
-\Delta\chi_1/\Delta\chi_2 = d_1/d_2 \approx 0.122$ so that the
dynamical phase difference between two successive positions of PS2 vanishes.
For an appropriate adjustment of the transmission
 coefficient, we use a gadolinium solution as absorber, which is tuned to exhibit
 a transmissivity of $T_{\text{abs}}=0.118(5)$. Taking the absorption of the $0.5$mm Al phase shifter
 into account, a ratio $T_{2}/T_1=0.120(5)$ is realized.

\begin{figure}[htbp]
  \centering
  \includegraphics[width=70mm]{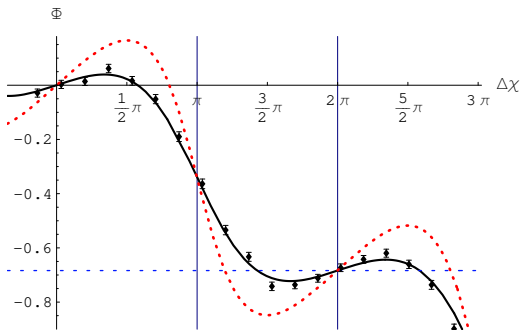}
  \caption{Observed phase shift $\Phi$ for a non-cyclic evolution of the
    state vector parameterized by the relative phase shift $\Delta\chi$. The dotted line indicates the
    theoretical prediction for the geometric phase assuming hundred percent visibility, whereas the
    solid line takes the diminished visibility into account. For a
    cyclic evolution ($\Delta\chi = 2\pi$) we obtain $\Phi_g = -0.684(48)$ radians for the transmission ratio $T_2/T_1 = 0.120(5)$.}
  \label{2004-splitbeam-prl-f3}
\end{figure}

 The phase shifts $\Phi$ of the sinusoidal intensity modulations due to PS1
 are determined at various points on the path traced out by the state,
 corresponding to a non-cyclic evolution. In practice, this is achieved by measuring
 the intensity modulation by PS1 at various positions of the PS2 \cite{hasegawa96}.
 The parameter of the evolution is the relative phase shift $\Delta\chi$, which was varied from $-0.2\pi$ to $3.0\pi$.
 The measured phase shift is plotted as the function of $\Delta\chi$
 in Fig. \ref{2004-splitbeam-prl-f3} together with theoretically predicted curves:
 one (dotted curve) is obtained by assuming an ideal situation of
 hundred percent visibility for all loops,
 whereas the practically diminished visibility is taken into account for the other (solid curve).
In particular, the two subbeams from the second loop are
only partially  overlapping (in space) with the reference beam at BS6
due to unequal spatial displacements caused by the unequal thicknesses
of the plates of PS2. These non-overlapping parts do not contribute to
the interference pattern that in turn induces a flattening of the
measured curve relative to the ideal curve. 
Other, however minor, contributions to this flattening are
from inhomogenous phase
distributions and transmission coefficients leading to an incoherent
superposition of states. Averaging over such a state distribution gives rise to
additional damping terms $e^{-\Gamma_i}$ in each beam, i.~e. $\Phi'=
\arg[\sqrt{T_1}e^{-\Gamma_1}e^{i\chi_1} +
\sqrt{T_2}e^{-\Gamma_2}e^{i\chi_2}]$ in contrast to $\Phi$ in
Eq. (\ref{eq:phase}), which can also be explained in terms of a mixed
state geometric phase \cite{sjoqvist00}.

All the mentioned influences are subsumed in the 
fit coefficient $C=0.57(2)$ obtained from a least-squares fit (solid line)
to the measured data using the function
 $\arg[\sqrt{T_1}e^{-i s_1\Delta\chi} + C
 \sqrt{T_2}e^{i s_2\Delta\chi}]$ with
 $s_{1,2}=d_{1,2}/(d_1+d_2)$, 
\footnote{The terms $s_{1,2}$ are due to $\chi_{1,2} \propto \mp d_{1,2} \gamma
  = \mp  \Delta\chi d_{1,2}/(d_1+d_2)$ with $\gamma = \Delta\chi/(d_1+d_2)$.}
which is a version of Eq. (\ref{eq:phase}) adapted
 to the experimental situation. Note, that these experimental factors
do not invalidate the discussion on the vanishing dynamical
phase: The deviation of the experimentally determined (solid)
curve from the ideal (dotted) curve is due to the measurement
circumstances in the neutron interferometer. The remaining contribution of the
dynamical phase due to the slightly different ratios of
$-\Delta\chi_1/\Delta\chi_2$ and $T_2/T_1$ can be calculated to yield
$\Phi_d = -0.009(26)$ at $\Delta\chi=2\pi$.

One can recognize the increase of the measured phase shift
$\Phi$
 in Fig. \ref{2004-splitbeam-prl-f3} due to the positively
 oriented surface on the Bloch-sphere
 (c.f. Fig. \ref{2004-splitbeam-prl-f2b}) followed by a decrease due
 to the counter-clockwise transversed loop yielding a negative phase contribution. This behaviour
 clearly exhibits the geometric nature of the measured phase.
 For a cyclic evolution ($\Delta\chi = 2\pi$) the measured phase is
 $-0.684(48)$ which is in a good agreement with the analytical value
 $-0.683$ of the geometric phase for a ratio $-\chi_1/\chi_2 = T_2/T_1 = 0.5/4.1$.

Another indication for a measurement of a non-cyclic geometric phase
is the varying amplitude of the interference
fringes dependent on $\Delta\chi$\cite{wagh99}. However, for the
absorption ratio  $T_2/T_1 = 0.122$ these differences are  at the detection limit. 
Measurements of other $T$-values are of interest and detailed results
of such measurements will be published in a forthcoming publication.

In summary we have shown that one can ascribe a geometric phase
not only to spin evolutions of neutrons, but also to evolutions in
the spatial degrees of freedom of neutrons in an interferometric
setup. This equivalence is evident from the description of both
cases via state vectors in a two dimensional Hilbert space.
However, there have been arguments contra the experimental
verification in \cite{hasegawa96} which we believe can be settled
in favour of a geometric phase appearing in the setup described
above. The twofold calculations of the geometric phase either in terms
of a shift in the interference fringes or via surface integrals in
an abstract state space allows for a geometric interpretation of
the obtained phase shift. The experiments exhibit a shift of the
interference pattern that reflects these theoretical predictions
up to influences due to the different visibilities in
the different beams.

This research has been supported by the Austrian Science
Foundation (FWF), Project Nr. F1513. S. F. wants to thank
E. Sj\"oqvist  for valuable discussions and K. Durstberger for critical
readings of the manuscript.

\bibliographystyle{apsrev}

\end{document}